\documentclass[prb,aps,twocolumn,showpacs,amsmath,amssymb]{revtex4-1}

\usepackage{graphicx}
\usepackage{psfrag}
\usepackage{epsfig}
\usepackage{color}
\usepackage[tight]{subfigure}
\definecolor{dred}{rgb}{0.7,0.0,0.0}
\usepackage{dcolumn}
\usepackage{bm}
\allowdisplaybreaks 
\usepackage{braket}
\usepackage{hyperref}
\hypersetup{colorlinks=true, linkcolor=blue, citecolor=blue, filecolor=blue, urlcolor=blue, pdftitle=, pdfauthor=, pdfsubject=, pdfkeywords=}

\newcommand{\cnod}{c^{\phantom{\dagger}}}

\begin{document}
\title{Exact-diagonalization results for resonant inelastic x-ray scattering spectra of one-dimensional Mott insulators}
\author{Stefanos Kourtis}
\author{Jeroen van den Brink}
\author{Maria Daghofer}
\affiliation{Institute for Theoretical Solid State Physics, IFW Dresden, 01171 Dresden, Germany}
\date{\today}

\begin{abstract}
We examine the momentum-dependent excitation spectra of
  indirect as well as direct resonant inelastic x-ray scattering
  (RIXS) processes in half-filled (extended) Hubbard rings. We
  determine the fundamental features of the groundstate RIXS response
  and discuss the experimental conditions that can allow for the
  low-energy part of these features to be distinguished in
  one-dimensional copper-oxide materials, focusing particularly on the
  different magnetic excitations occurring in indirect and direct RIXS
  processes. We study the dependence of spin and charge excitations
on the choice of and detuning from resonance. Moreover, final state
excitation weights are calculated as a function of the core-hole
potential strength and lifetime. We show that these results can be
used to determine material characteristics, such as the core-hole
properties, from RIXS measurements.  
\end{abstract}

\maketitle

\section{Introduction} \label{sec:intro}

Strongly interacting one-dimensional (1D) quantum systems provide the
opportunity to observe distinctive signatures of theoretical
predictions, such as the separation of fermionic excitations into
charge and spin components.~\cite{Lieb1968} Prototypical materials
which exhibit 1D behavior are the corner-sharing copper-oxide
compounds Sr${}_2$CuO${}_3$ and SrCuO${}_2$, in which 3D
antiferromagnetic correlations disappear above a temperature $T_N$ of
a few Kelvin and the majority of electronic properties comes from
electrons moving on almost independent chains. A variety of
experimental techniques has been used to verify the 1D character of
the excitations in these
materials~\cite{Motoyama1996,Matsuda1997,Keren1993,Kojima1996,Kojima1997}
and the existence of the expected spin-charge
separation.~\cite{Kim1996,Kim1997,Neudert1998,Kim2006} These observations have
induced further interest in cross-examining the results of different
experimental probes, in order to test the quantitative agreement with
theoretical descriptions of the underlying quantum mechanical
processes. 

Resonant inelastic x-ray scattering (RIXS) is a particularly promising
technique for research in this direction, since it can probe multiple
spin and charge excitations on the same footing.~\cite{Ament2011} In RIXS, incident x-ray
photons are used that have a frequency close to the
corresponding excitation energy for the promotion of a core electron
to an empty state in or above the valence orbital of interest, the two
alternatives defining direct and indirect RIXS
respectively.~\cite{VANDENBRINK2005,Brink2006} In the intermediate
state the resulting core hole perturbs the valence electron system locally, until
an electron decays into the empty core state, emitting an x-ray photon
shifted in energy, momentum and polarization with respect to the
incoming one. RIXS experiments at the Cu $K$-edge have been performed
on the aforementioned 1D
cuprates~\cite{Hasan2002,Kim2004a,Qian2005,Suga2005,Seo2006} and the
dominant spectral features have been attributed to $d$-$d$
excitations. Spin excitations were not distinguished in the earlier
experiments, because of resolution limitations of the instrumentation
used. These limitations can be overcome in today's experimental
setups, as has been demonstrated by the observation of spin
excitations in other 1D~\cite{Schlappa2009} or
2D~\cite{Hill2008,Braicovich2010} cuprates and iridates,~\cite{Kim2011,Ament2011,Veenendaal2011} although RIXS
measurements for the simpler Sr-based chain cuprates are still awaited.~\cite{Schlappapre2011}

The distinct form of the theoretical expression for the RIXS cross
section provides a unique way to study properties of models for
one-dimensional condensed matter systems. The shortness of the
intermediate state lifetime has been exploited to perform expansions
for the magnetic indirect RIXS cross section of Mott
insulators.~\cite{VandenBrink2007,Ament2007,Nagao2007a} 
The case of magnetic excitations in direct RIXS has been investigated 
in this manner~\cite{Ament2009} and also by an effective operator method,~\cite{Haverkort2010} recently applied to the case of the quasi-1D compound TiOCl.~\cite{Glawion2011} Apart from
model fits to experimental data,~\cite{Kim2004a,Qian2005} exact
numerical investigations of indirect RIXS on small systems have been
performed for the 1D (extended) Hubbard,~\cite{Tsutsui2000}
Heisenberg,~\cite{Klauser2011a,Forte2011} $t$-$J$~\cite{Forte2011} and $dp$~\cite{Okada2006}
models, whereas only a calculation of direct RIXS spectra for the Heisenberg and $t$-$J$ models has appeared so far.~\cite{Forte2011}
Furthermore, in the existing works, only the momentum
dependence of RIXS spectra at specific resonance conditions has been
obtained. It has been established experimentally that the character of
excitations present in the final state of RIXS depends strongly on the
detuning from resonance. Therefore, it is of interest to determine the
effect of detuning on final state excitations in a generic model and
see how it compares to the behavior of real systems. Moreover, the
effect of the intermediate-state core hole, which will play an
increasingly important role with the advent of time-resolved RIXS
experiments, has been considered only heuristically so far. 

The purpose of this paper is to present a systematic numerical
study of both indirect and direct RIXS processes in a 1D
Mott-insulating system. To this end, the Lanczos exact diagonalization
method is used to calculate the RIXS spectra for the Hubbard and
extended Hubbard models at half filling. The focus is on determining
fundamental properties and extracting trends of the models under changes in the core-hole parameters, rather
than fitting experimental data. However, as the (extended) Hubbard model
is believed to capture the essential physics of 1D chain cuprates at
low energies, our results should be adequate to make
qualitative predictions for the experimental outcome in this energy
range. In Sec.~\ref{sec:rixs_theory} the theoretical formulation
leading to the evaluated RIXS cross section is briefly reviewed. In
Sec.~\ref{sec:results} the RIXS response is calculated as a function
of incoming photon energy. Moreover, the dependence of the RIXS
spectra upon the values of core-hole potential and lifetime is
carefully investigated and it is shown how the core-hole parameters
can be related to direct RIXS experiments. Our conclusions are
summarized in Sec.~\ref{sec:conclusions}. 

\begin{figure}[b]
\includegraphics[width=0.85\columnwidth]{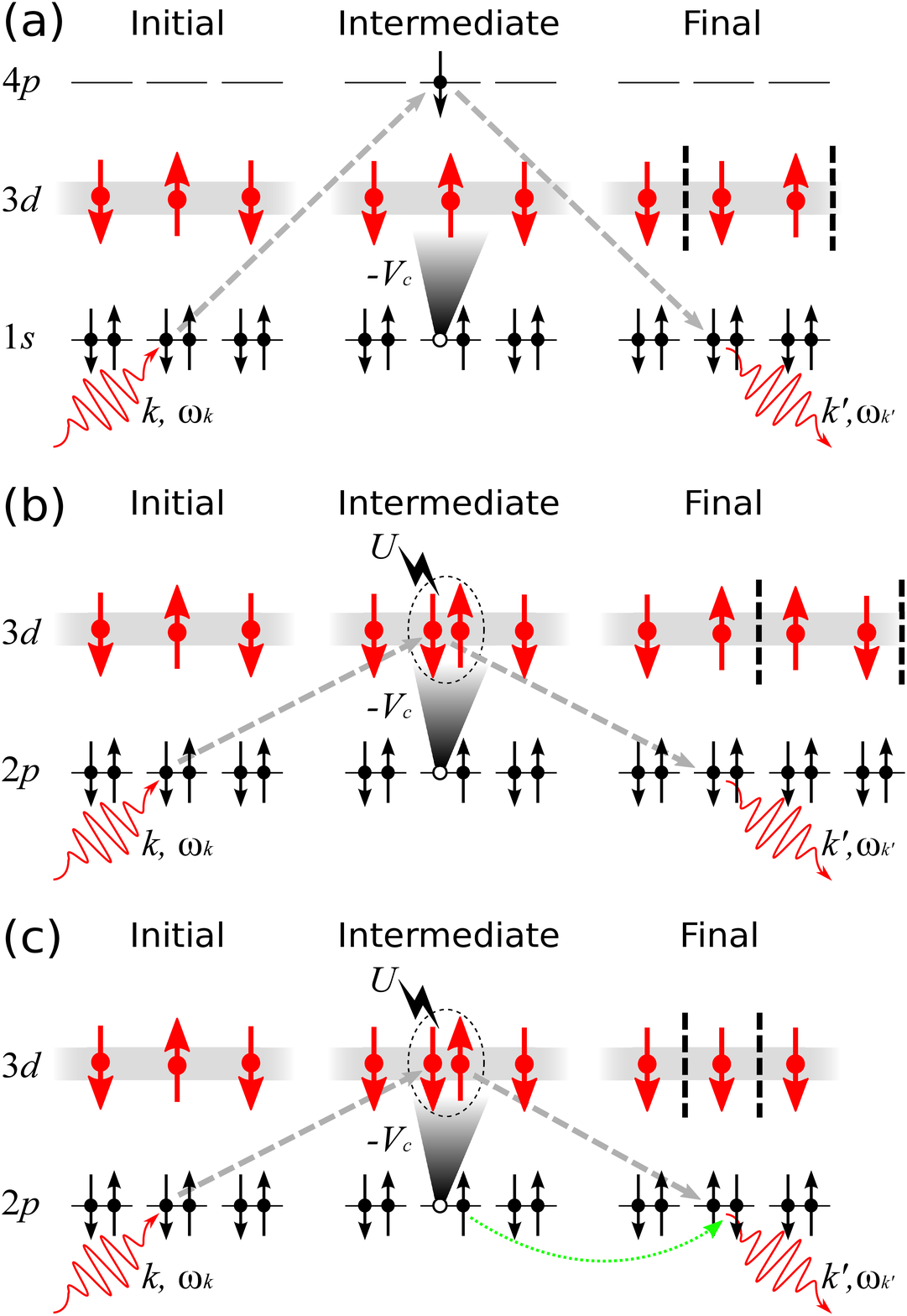}
\caption{(Color online) Illustration of RIXS processes in half-filled copper oxides. (a) Indirect RIXS process. (b) direct RIXS process without spin flip. (c) direct RIXS process with spin flip. For a spin flip to occur, the core-hole spin must also be flipped (dotted arrow).}
\label{fig:rixs}
\end{figure}

\section{RIXS cross section} \label{sec:rixs_theory}

\subsection{Kramers - Heisenberg formula}

The unperturbed Hamiltonian $\hat H_0$ of choice for the $N$-site, 1D system we wish to study is the single-band Hubbard model:
\begin{equation}
\hat {H_0} = - t \sum_{j\sigma}^N (c_{j\sigma}^\dagger c_{j+1\sigma}^{\phantom{\dagger}} + \textrm{h.c.}) + U \sum_{j}^N \hat n_{j\uparrow} \hat n_{j\downarrow} ,
\end{equation}
where $c_{j\sigma}^\dagger$ creates electrons with spin $\sigma$ at
site $j$, $\hat n_{j\sigma} = c_{j\sigma}^\dagger \cnod_{j\sigma}$,
$t$ is the hopping integral and $U$ the on-site Coulomb repulsion
strength. In one dimension, any non-zero $U/t$ ratio is enough to
guarantee a Mott-insulator ground state with energy $E_g$ at half
filling.~\cite{Lieb1968} 
For comparison, we also study the extended Hubbard model with an additional
nearest-neighbor interaction
\begin{align}
\hat{H}_{\textrm{ext}} = \hat {H_0} + V\sum_{j\sigma\sigma'}^N\hat{n}_{j\sigma}\hat{n}_{j+1\sigma'}
\end{align}
with $V=2t$. 
The core hole present in the intermediate
state acts as a localized potential scatterer for the $3d$ electrons
and can be expressed as 
\begin{equation}
 \hat H_{\textrm{int}} = - V_c \sum_{j\sigma\sigma'}^N \hat n^{\phantom{h}}_{j\sigma} \hat n^h_{j\sigma'},
\end{equation}
where $\hat n^h_{j\sigma}$ is the number operator for core holes,
$\hat n^h_{j\sigma} =1$ signifying the presence of a core hole with
spin $\sigma$ on site $j$. The core hole interacts via an attractive
force parameterized by $V_c > 0$ with the total density 
$\hat n_{j}=\hat n_{j\uparrow} + \hat n_{j\downarrow}$ in the valence
band described by $\hat {H_0}$. The dipole operator that creates
the intermediate state, as schematically illustrated in
Fig.~\ref{fig:rixs}, is  given by
\begin{equation}
 \hat D = \sum_{j\sigma}^N e^{i k R_j} d_{j\sigma}^\dagger h^{\phantom{\dagger}}_{j\sigma},
\end{equation}
where $h_{j\sigma}$ creates a core hole and $d_{j\sigma}^\dagger$ a
photoexcited electron at position $R_j$; $k$ is the incoming photon
momentum, corresponding to frequency $\omega_{k}$. Here we do not consider photon
polarizations and geometrical factors. The RIXS cross
section is given by the Kramers - Heisenberg
formula:~\cite{Kramers1925} 
\begin{equation}
w \propto \sum_f |F_{fg}| ^2 \delta(E_f-E_g-\Delta\omega),\label{rixs_cross}
\end{equation}
where the sum is over all final states $\ket{f}$ with energies $E_f$. The RIXS amplitude has been defined as
\begin{equation}
 F_{fg}(\omega_{\textrm{in}}) \equiv \sum_n \frac{\braket{f | \hat
     D'^\dagger | n}\braket{n | \hat D | g}}{E_g - E_n +
   \omega_{\textrm{in}} + i\Gamma}, 
\end{equation}
where $\omega_{\textrm{in}}\equiv\omega_{k}-\omega_{\textrm{res}}$ is
the detuning of the incoming photon frequency from resonance and $E_n$
and $\Gamma$ are the energy and lifetime of the intermediate
states. The fact that core holes are usually localized allows us to
rewrite the above expression as 
\begin{equation}
 F_{fg}(\omega_{\textrm{in}}) = \sum_{j\sigma\sigma'}^N e^{i q R_j}
 \braket{f | d_{j\sigma'} \frac{1}{\hat {H_j} - E_g -
     \omega_{\textrm{in}} - i \Gamma} d_{j\sigma}^\dagger |
   g},\label{rixs_scamp} 
\end{equation}
where $q \equiv k - k'$ is the momentum transfer and $\hat {H_j}
\equiv \hat {H_0} - V_c \hat n_j$ is a locally perturbed
Hamiltonian. 
For indirect RIXS, one has $d_{j\sigma}^\dagger\equiv 1$,
i.e., the only impact of the photon is the creation of the core
hole, as illustrated in Fig.~\ref{fig:rixs}(a). For direct 
RIXS $d_{j\sigma}^\dagger\equiv c_{j\sigma}^\dagger$ photo-excites an
electron into the band modelled by the (extended) Hubbard 
model, see Fig.~\ref{fig:rixs}(b,c). In the case of direct RIXS, care must be taken to sum over both
the RIXS processes involving a photo-excited up electron and the the one
involving a down electron, because the singlet character of the ground
state is otherwise not preserved.

\subsection{Method}


We use the Lanczos method~\cite{Lanczos1950,Lanczos1952} to obtain the
ground state of finite Hubbard rings and then to evaluate the RIXS
cross section of Eq.~\eqref{rixs_cross}. Even though the RIXS final
state is translationally invariant, the intermediate-state Hamiltonian
$\hat {H_j}$, which is needed to evaluate the Green's functions acting
on the initial state, is not. Translational invariance can thus not be
used to reduce the Fock space size and all results presented here
are for 14-site rings. RIXS 
spectra of smaller even-site, as well as of 16-site rings have also
been selectively calculated and were found to be qualitatively and
quantitatively consistent with the 14-site results. 

Related numerical calculations of RIXS spectra using a similar method
have been carried out~\cite{Tsutsui2000} for a few specific cases and
the results are in agreement with ours.  We use $t$ as unit of energy,
the on-site Coulomb repulsion
is kept fixed at $U=10t$, and where the extended Hubbard model is
studied, we set $V=2t$.  We have used Lanczos exact
diagonalization to explore 
signature features of the RIXS response and its dependence on the
remaining parameters of the Hamiltonian describing the core hole
potential and the energy of the incoming photon.

\section{Results} \label{sec:results}

\subsection{Spin excitations in RIXS spectra}

\begin{figure}
\includegraphics[width=0.8\columnwidth]{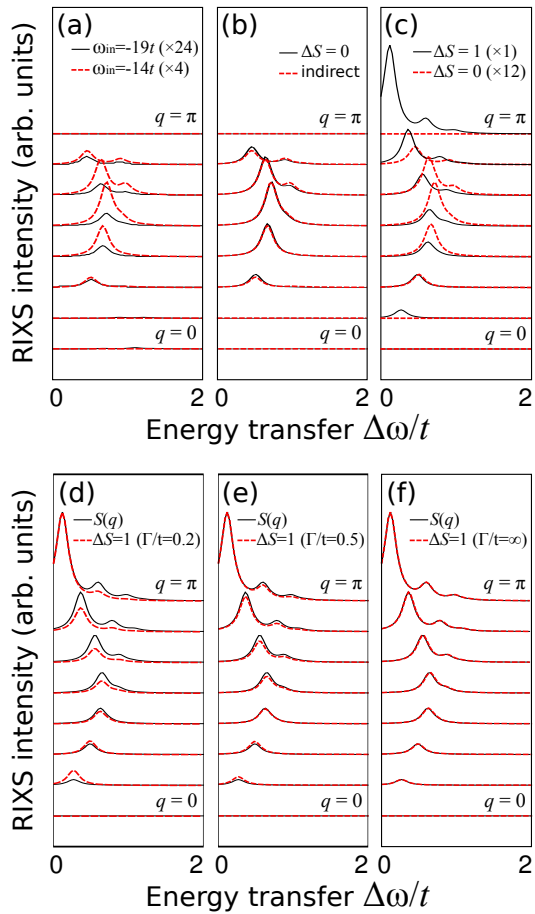}
\caption{(Color online) Overview of magnetic excitations in calculated RIXS spectra
  for the Hubbard model on a half-filled, 14-site ring. The parameters are $U/t=10$, $V_c/t=15$,
  $\omega_{\textrm{in}}/t=-19$, $\Gamma/t=1$, unless otherwise
  noted. The peaks are Lorentzians of width $0.1t$. Note that
  RIXS spectra in (a) and (c) are rescaled by the factors in parentheses. First row: (a) Indirect RIXS spectra for input energies corresponding
  to well-screened (solid line) and poorly screened (dashed line)
  intermediate states. (b) $\Delta S=0$ direct RIXS spectrum (solid
  line) compared to indirect RIXS (dashed line). (c) $\Delta S=1$
  direct RIXS spectrum (solid line) compared to $\Delta S=0$ direct
  RIXS (dashed line). Second row: Spin DSF for the same system (solid line)
  compared to $\Delta S=1$ direct RIXS spectrum (dashed line) for (d) $\Gamma/t=0.2$, (e) $\Gamma/t=0.5$, and
  (f) ``instant'' ($\Gamma/t=\infty$) RIXS process.} 
\label{fig:rixs_spin}
\end{figure}

The outcome of an indirect RIXS process depends strongly on the
resonance condition. Calculation of the x-ray absorption (XAS)
spectrum for the Hubbard model shows that there are two main
resonances for indirect RIXS:~\cite{Tsutsui1999} one corresponds to
intermediate states where the valence band at the core-hole site is doubly occupied and
the core hole is therefore well-screened, at energies $\sim U -
2V_c$ with respect to the initial state, and the other to states where
the core-hole site is singly occupied and the core hole is poorly
screened, at energies $\sim -V_c$ with respect to the initial
state. In states with double occupancies at the core-hole sites, 
exchange interactions are not favorable and therefore spin excitations
are not pronounced, whereas they are favored when core-hole sites are
singly occupied. This is evident in 
Fig.~\ref{fig:rixs_spin}(a), where the low-energy parts of the
indirect RIXS spectra taken close to the two resonances are presented.

In direct RIXS on the other hand, a double occupancy is created at the
core-hole site by default at half filling and therefore there is only one resonance,
corresponding to well-screened intermediate states. The form of the
RIXS scattering amplitude \eqref{rixs_scamp} suggests that in direct
RIXS there can be processes where $\sigma=\sigma'$ ($\Delta S=0$) or
$\sigma\not=\sigma'$ ($\Delta S=1$). In a direct RIXS experiment both
processes take place. For a $\Delta S=1$ process to occur however, a
spin-orbit interaction between the spin and angular degrees of freedom
of the core hole is necessary (see Fig.~\ref{fig:rixs}). Hence, the
weight ratio of the two direct RIXS processes depends heavily on the
strength of the core-hole spin-orbit coupling. 

The low-energy parts of the spectra for the two direct RIXS processes
are depicted in Figs.~\ref{fig:rixs_spin}(b,c). It is seen that
excitations in the low-energy part of the $\Delta S=0$ process
spectrum are almost identical to those occurring in indirect RIXS
[Fig.~\ref{fig:rixs_spin}(a)], although in this case they are caused
by the decoupling of the intermediate state doublon into an antiholon
and a spinon. The $\Delta S=1$ process spectrum is very similar to the
spin dynamical structure factor (DSF) $S(q)$ for the studied system,
shown in Figs.~\ref{fig:rixs_spin}(d-f), and corresponds mainly to
single-magnon excitations. 
Indeed, by setting
$\Gamma=1/\tau\rightarrow\infty$, i.e., by assuming an instantaneous
RIXS process with an infinitely fast decay of the core hole, $S(q)$ is
obtained exactly, see Fig.~\ref{fig:rixs_spin}(f). 
The differences between $S(q)$ and the
$\Delta S=1$ direct RIXS spectrum in  Figs.~\ref{fig:rixs_spin}(d,e) are
thus a direct consequence of the core hole's finite lifetime in the intermediate state.

The aforementioned differences between $S(q)$ and the $\Delta S=1$ direct RIXS spectrum consist of changes in the weight distribution along the Brillouin zone, while the total weight remains constant, as will be shown below. Even though the $\Delta S=1$ direct RIXS spectrum may appear to be almost insensitive to the value of $\Gamma$, the effect of a finite core-hole lifetime is actually important for the determination of the lineshape, even though the main excitation in a $\Delta S=1$ direct RIXS process - namely the single magnon - does not depend on the presence of the core hole. On the contrary, indirect and $\Delta S=1$ direct RIXS processes depend crucially on core-hole properties (see discussion below).
%

The generic features of the spectra in Fig.~\ref{fig:rixs_spin} are similar to those found in linear spin-wave theory results,~\cite{Haverkort2010} but they have more structure. For example, in $\Delta S = 1$ direct RIXS several peaks arise when moving towards the Brillouin zone boundary, apart from the single peak of single-magnon excitations. Also, the two-magnon spectral function is found to be finite at the zone boundary using spin-wave theory, but in Figs.~\ref{fig:rixs_spin}(b,c) we see that the $\Delta S = 0$ direct RIXS amplitude vanishes there. This means that two-magnon excitations at $q=\pi$ can arise only in $\Delta S = 1$ direct RIXS. Our results are also consistent with numerical Bethe ansatz calculations of dynamical spin and spin-exchange structure factors for the Heisenberg model,~\cite{Klauser2011a} which are the zero-lifetime limits of $\Delta S = 1$ direct and indirect RIXS processes respectively.

The $\Delta S=1$ part of the direct RIXS response is one order of
magnitude larger than that of the $\Delta S=0$ process. This is
because in the $\Delta S=1$ case there will be a magnon excitation in
the final state even if no scattering with the core-hole potential
occurs in the intermediate state. In the $\Delta S=0$ case, at least
two scatterings of electrons off the core-hole potential are needed
for magnetic excitations to arise. In direct RIXS, the incoming
photon's angular momentum couples to the core electron's angular
momentum, which is in turn coupled to the electron spin via
spin-orbit coupling. This implies that it is in principle possible to
alter the ratio of the two processes by appropriate choice of the
incoming beam polarization and thus observe experimentally the
respective change in the character of the spin excitations. An
interesting feature seen in Fig.~\ref{fig:rixs_spin}(c) is the
difference between the one-magnon and the two-magnon peak
locations. This difference decreases with increasing chain length and
becomes zero for long chains. The fact that the two RIXS responses
scale differently with increasing system size is related to the
different size of the excitations present in each case. It is
noteworthy that experimental characterization of another quasi-1D cuprate,
CaCu${}_2$O${}_3$, suggests the presence of weakly coupled,
finite-size chain segments, 13-14 copper sites in
length.~\cite{Kiryukhin2001} Such an energy difference might thus
reveal signatures of chain breaks, in addition to an analysis of the static
magnetic susceptibility.~\cite{PhysRevLett.98.137205} 
If this is indeed the case, then our
results show that this energy difference can be resolved in direct
RIXS experiments and can possibly be used to distinguish the different
spin-excitation weights. Before proceeding, we note that all the
results presented in Fig.~\ref{fig:rixs_spin} are qualitatively the
same and quantitatively very similar for the extended Hubbard model,
for values of up to $2t$ for the nearest-neighbor Coulomb repulsion.  

\begin{figure}
\includegraphics[width=\columnwidth]{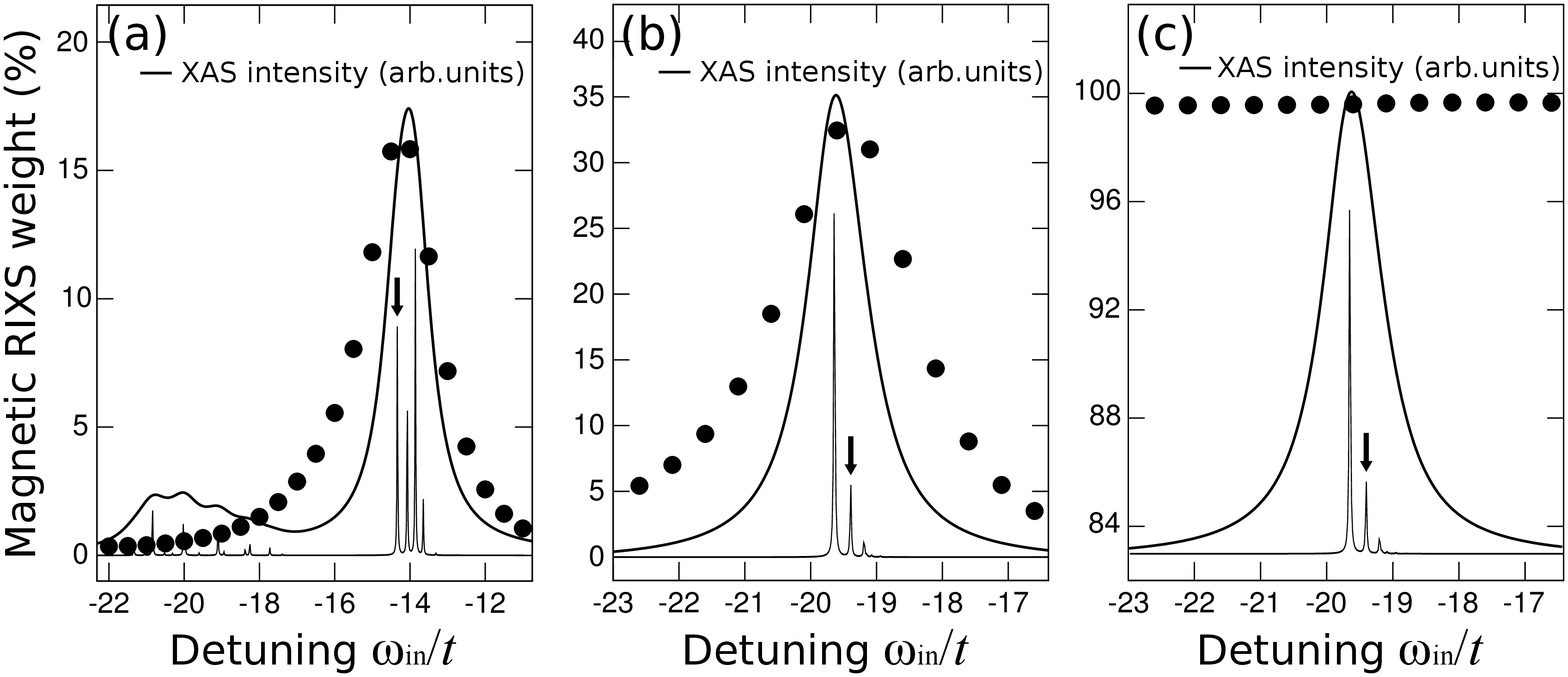}
\caption{Total spin excitation weight, summed over all inequivalent momenta $q$ in the first Brillouin zone, as a function of $\omega_{\textrm{in}}$ in (a) indirect, (b) $\Delta S = 0$  and (c) $\Delta S = 1$ direct RIXS respectively, for the same system as in Fig.~\ref{fig:rixs_spin}. The thick (thin) solid lines are the corresponding XAS spectra with $\Gamma_{\textrm{XAS}}$ set to $0.5t$ ($0.01t$). The arrows indicate the XAS peaks at which the magnetic RIXS signal is maximal.}
\label{fig:weights_rixs_nocharge}
\end{figure}

\subsection{Resonance condition and detuning}

The type of excitations present in the final state of RIXS is defined
not only by the type of transition in the absorption step, but also by
the position relative to the main resonance peak. By appropriately
tuning $\omega_{\textrm{in}}$, specific excitations can thus be
favored or hindered. This is evident for example in the dependence of
the total spin-excitation weight on $\omega_{\textrm{in}}$ around the
XAS peaks, as presented in Figs.~\ref{fig:weights_rixs_nocharge}(a) and
(b). The momentum dependence of the charge sector excitations can be
explained within a free particle-hole (doublon-hole) excitation
picture and is very similar to the one obtained in electron
energy-loss spectroscopy (EELS) measurements.~\cite{Neudert1998}
The energy-loss range of charge excitations in RIXS
  becomes narrower with increasing momentum and the main peak shifts
  to higher energy loss.~\cite{Tsutsui2000} Similar spectral shifts of
  charge excitation peaks are observed in numerical results for
  two-dimensional cuprates.~\cite{Tsutsui1999,Jia2011} A 
finite nearest-neighbor Coulomb repulsion $V$ is equivalent to an
attraction between neighboring doublon-hole pairs and leads to a
transfer of charge spectral weight to lower energies for all
momenta.~\cite{Tsutsui2000} Apart from this shift, the results
presented in Fig.~\ref{fig:weights_rixs_nocharge} remain qualitatively
unchanged for values of up to $V=2t$. 

\begin{figure}
\includegraphics[width=\columnwidth]{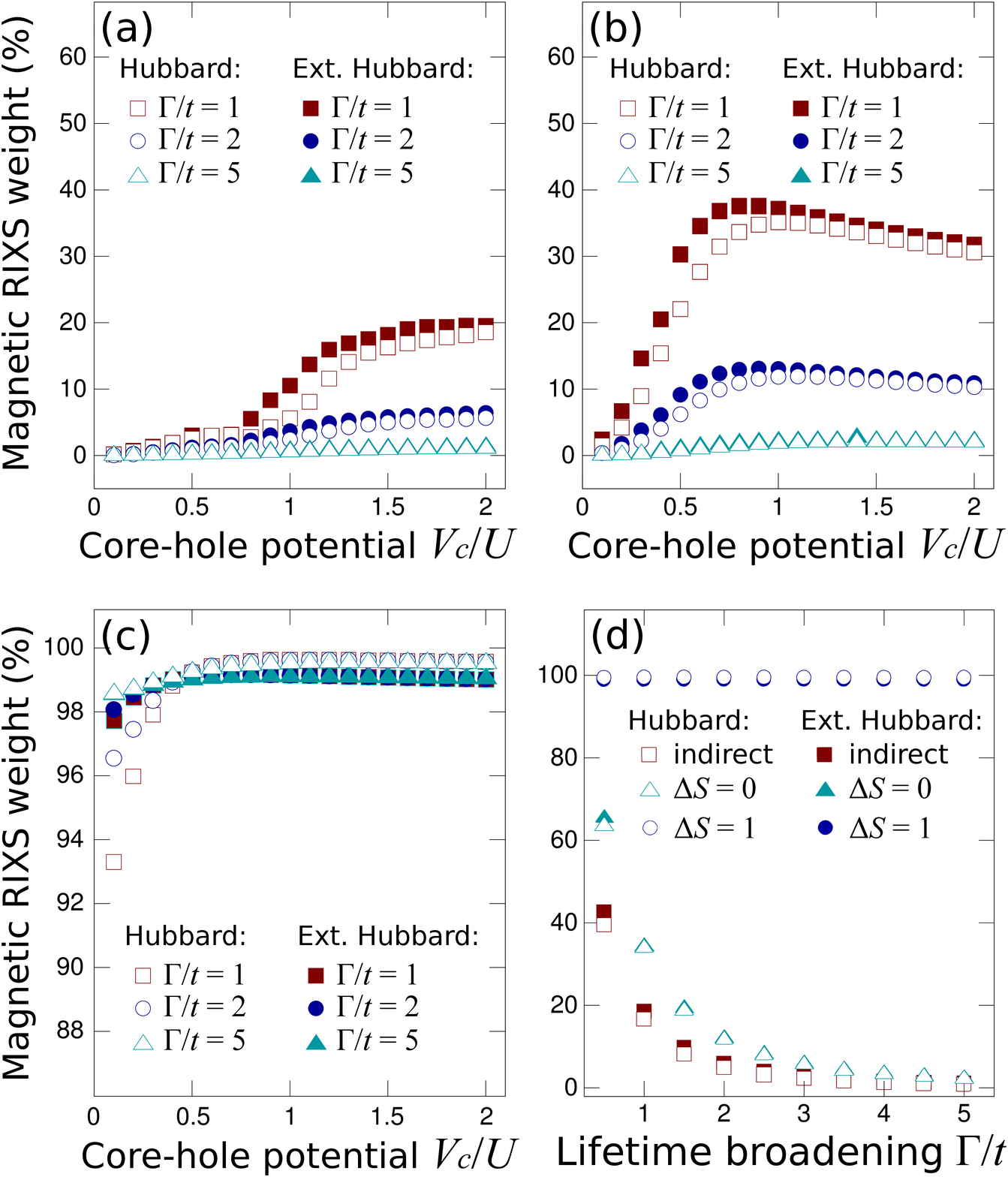}
\caption{(Color online) Total spin excitation weight percentage as a function of the core-hole potential in (a) indirect RIXS, (b) $\Delta S = 0$ direct RIXS and (b) $\Delta S = 1$ direct RIXS for Hubbard (empty symbols) and extended Hubbard (full symbols) models, for three values of the inverse lifetime $\Gamma$. The value for the detuning $\omega_{\textrm{in}}$ of the incoming photon energy from resonance at each point is defined by the procedure outlined in the text. All other relevant parameters are as in Fig.~\ref{fig:rixs_spin}. (d) Core-hole lifetime dependence of magnetic excitations in all RIXS processes for $V_c=15t$.}
\label{fig:corehole_pot}
\end{figure}

It is noted that the maxima of absorption in Figs.~\ref{fig:weights_rixs_nocharge}(a) and (b) do not coincide exactly with the maxima of spin excitations. Upon increasing the XAS spectrum resolution, it is seen that the maximal RIXS spin excitation response lies close to specific states within the broad peaks. Therefore, in order to study the dependence of excitations on the core-hole potential strength, $\omega_{\textrm{in}}$ must be carefully selected in each case. For this purpose, the $\omega_{\textrm{in}}$-dependence of the spectrum is investigated first for each value of $V_c$ and the value that yields the largest spin excitation weight is chosen as a reference. For example, the arrows in Fig.~\ref{fig:weights_rixs_nocharge} point to the appropriate $\omega_{\textrm{in}}$ values for the case of $V_c=15t$. After determining the reference states, the $V_c$-dependence of the total weights can be studied. The scaling of the total magnetic excitation weight, summed over all inequivalent momenta in the first Brillouin zone, is shown in Fig.~\ref{fig:corehole_pot} for indirect and direct RIXS processes.

\subsection{The role of the core-hole potential magnitude}

Figure~\ref{fig:corehole_pot} shows the relative spin excitation spectral weight as a function of the core-hole potential for three different core-hole lifetime values. As was mentioned above, the experimental direct RIXS spectra contain both the $\Delta S = 0$ and $\Delta S = 1$ contributions. It is in principle possible to deduce the contribution of each of the two processes by comparing the momentum-dependence of the experimental spectrum lineshape (with all geometrical factors stripped off) to the lineshapes of Figs.~\ref{fig:rixs_spin}(b,c). One can then compare with the results presented in Fig.~\ref{fig:corehole_pot} and determine possible values for the core-hole potential and lifetime, which are typically considered as fit parameters. We note that this estimation scheme is general and therefore applicable to more complicated models, provided that the spin excitations can be clearly distinguished from other excitations, like e.g. $d$-$d$ excitations in multiple-orbital models.

The spin-excitation weight for indirect RIXS in
Fig.~\ref{fig:corehole_pot}(a) shows a dip around $V_c=U$ for both
Hubbard and extended Hubbard models. This can be intuitively
understood as follows: for these values of the core-hole potential,
the attraction exerted by the core hole to an electron on a different
site is approximately canceled by the effect of the
on-site Coulomb repulsion due to the electron already occupying the
site with the core hole. There is therefore no cause for scattering
and spin exchange processes are consequently suppressed. We find no
such dip in $\Delta S = 0$ direct RIXS, a fact which demonstrates the
different origin of the same excitations in the two RIXS processes. As
can be seen in Fig.~\ref{fig:corehole_pot}, the inclusion of a
nearest-neighbor Coulomb repulsion with $V/t = 2$ does not
substantially alter spin-excitation weights. 

\subsection{RIXS core-hole lifetime}

It has been noticed in previous studies that the value of the
intermediate state lifetime $1/\Gamma$ affects the weight of spin
excitations in model calculations.~\cite{Okada2006} The reason is that
spin exchange processes are slow compared to charge excitations. This
is indeed the case for indirect RIXS. In direct RIXS however, the
single-spin flip is, as discussed above, a zero-order process in terms
of electron -- core -hole scattering, and the total
excitation weight does  therefore not depend on $\omega_{\textrm{in}}$, $\Gamma$
or $V_c$, even though the lineshape depends on these parameters,
as is shown in Figs~\ref{fig:rixs_spin}(d-f) for the case of
$\Gamma$. This implies that in direct RIXS, evidence of single-spin
excitations should be experimentally observable in 1D chain materials,
as is the case at the Cu $L$ edge in 2D
cuprates~\cite{Hill2008,Braicovich2010} and in
Sr${}_{14}$Cu${}_{24}$O${}_{41}$.~\cite{Schlappa2009} Regarding the
double spin-flip processes in RIXS, the core-hole lifetime has been
shown to be sufficient to detect such spin excitations in the case of
the the Cu $K$ edge in 2D 
cuprates.~\cite{Hill2008,Bisogni2010} It is thus justified to assume that
the intermediate-state lifetime is similarly long enough in chain
systems, suggesting that spin excitations due to exchange interactions
should be present in experimental RIXS spectra. 

As can be seen in Fig.~\ref{fig:corehole_pot}(a-c), magnetic excitation weights remain unchanged for large
ranges of $V_c$, in both direct and indirect RIXS. Since it is very
likely that the physical value of 
the core-hole attraction is within these ranges, the core-hole
lifetime dependence of magnetic excitations shown in
Fig.~\ref{fig:corehole_pot}(d) can be used to identify the lifetime
range corresponding to experimental observations. The currently used
estimates based on the XAS core-hole lifetime, as well as experimental
RIXS evidence,~\cite{Bisognipre2011} suggest that $\Gamma/t<1$, a fact
that underlines the significance of the effects of the finite
core-hole lifetime presented above. This approach to
  obtain core-hole lifetimes is usable even in cases where the
  lifetime does not correspond to the natural line width. In the
  future, it may thus 
  become particularly useful for the interpretation of
  time-resolved RIXS measurements, in which the intermediate-state
  lifetime will be controlled via stimulated emission.

\section{Conclusions} \label{sec:conclusions}

We have determined the direct and indirect RIXS responses of the
half-filled, single-band Hubbard model in one dimension by using exact
diagonalization, systematically varying relevant parameters. In the
low-energy sector, our results exhibit the main magnetic excitations
allowed theoretically, namely two-magnon excitations in the indirect
and $\Delta S = 0$ direct RIXS processes and single-magnon excitations
in $\Delta S = 1$ processes. The strong dependence of the total spin
excitation weight on the resonance condition, as well as on the
detuning from resonance, is apparent in the RIXS spectra. The
characteristics of spin excitations in all RIXS processes remain
qualitatively unchanged by the introduction of a nearest-neighbor
Coulomb repulsion with magnitude of up to $2t$. The general features
of the calculated spectra are accessible to experiment and it is of
interest to see to what extent the obtained generic results can
describe measured magnetic features of 1D single-chain cuprates, which
have yet to appear. 

Furthermore, we have studied the dependence of spin excitations created by RIXS on the strength of the core-hole potential and we have presented a method for estimating the core-hole properties of interest using our or equivalent results, in conjunction with the respective geometrical factors. It is also seen that spin excitations in indirect RIXS are suppressed for values of $V_c$ close to the on-site Coulomb repulsion strength $U$ due to suppression of electron scattering off the core hole. The total weight of magnetic excitations in $\Delta S = 1$ direct RIXS does not depend on incoming photon energy, core-hole potential magnitude or core-hole lifetime, and can be used as a reference by which the total $\Delta S = 0$ contribution can be distinguished in experimental data. Using the ratio of the magnetic weights, the core-hole properties can be estimated.

During the writing of this manuscript, a preprint by Igarashi and Nagao appeared,~\cite{Igarashi2011} in which they apply their previously developed formalism~\cite{Igarashi2011a} to 1D systems and obtain similar results to the ones in Fig.~\ref{fig:rixs_spin}.

\begin{acknowledgments}
The authors are grateful to L.~Ament, V.~Bisogni and K.~Wohlfeld for invaluable discussions. This work was supported by the Deutsche Forschungsgemeinschaft under the Emmy-Noether program (SK and MD). 
\end{acknowledgments}

%

\end{document}